# Heat-driven spin transport in a ferromagnetic metal

Yadong Xu[1], Bowen Yang[1], Chi Tang[1], Zilong Jiang[1], Michael Schneider[2], Renu Whig[2], and Jing Shi[1]

1. Department of Physics & Astronomy, University of California, Riverside, CA 92521
2. Everspin Technologies, Chandler, AZ 85224

As a non-magnetic heavy metal is attached to a ferromagnet, a vertically flowing heat-driven spin current is converted to a transverse electric voltage, which is known as the longitudinal spin Seebeck effect (SSE). If the ferromagnet is a metal, this voltage is also accompanied by voltages from two other sources, i.e. the anomalous Nernst effect in both the ferromagnet and the proximity-induced ferromagnetic boundary layer. By properly identifying and carefully separating those different effects, we find that in this pure spin current circuit the additional spin current drawn by the heavy metal generates another significant voltage by the ferromagnetic metal itself which should be present in all relevant experiments.

The longitudinal spin Seebeck effect (SSE), a pure spin current effect driven by a vertical heat current, was established in the bilayer structures consisting of a magnetic insulator such as yttrium iron garnet and a heavy metal such as Au with strong spin-orbit coupling (SOC) but a negligible magnetic proximity effect [1, 2]. In conducting ferromagnets, the longitudinal SSE phenomenon is accompanied by two other sources: the anomalous Nernst effect (ANE) in both the ferromagnetic layer such as NiFe and the proximity-induced boundary layer in the non-magnetic material such as Pt. As pointed out by Huang et al. [3, 4], these two effects can overwhelm the SSE signal which was first reported in NiFe [5]. In metallic structures, separating the pure spin current effect from ANE generated by the ferromagnetic metal as well as proximity-induced ANE in the non-magnetic metal remains a serious challenge. In this work, we attempt to isolate these effects and quantify the relative contributions in metallic structures.

SSE is an outcome of spin diffusion occurred in heavy metals (e.g. Pt or Au) with strong SOC on the characteristic length scale called the spin diffusion length, $\lambda_S$; therefore, the magnitude of the effect depends on the dimension of the metal along the spin diffusion. ANE is usually treated as a property of bulk ferromagnetic materials [6]. In this study, we demonstrate that when the relevant dimension in a ferromagnet is comparable with its own $\lambda_S$, the spin current effect cannot be ignored. It produces an additional inverse spin Hall effect (ISHE) signal which is superimposed on the ANE signal of the ferromagnet. Analogous to electric circuits, the spin current flow in the pure spin current circuits can be effectively controlled by adjusting its source (temperature gradient in ferromagnet), drain (heavy metals), and channel (light metals).

The inset of Fig.1 shows the device geometry we use for ANE measurements. We first prepare a photoresist mask on a GaAs substrate that defines the channel (0.2 mm wide and 3.8 mm long) and two side electrodes and then deposit the metal film by sputtering. After covering portion of the side electrodes, we deposit a 300 nm thick $Al_2O_3$ layer followed by a 50 nm thick Au heater layer. The $Al_2O_3$ layer electrically insulates the heater from the sample beneath. Finally, the photoresist is lifted off with PG remover and the device with a self-aligned heater on top is resulted. This device structure generates a more uniform and controllable vertical temperature gradient, $\nabla T$, which is crucial to our study. Variations in deposition conditions can result in small but discernable variations in the voltage signal even with the same heater power. To avoid the source of this irreproducibility and therefore to resolve small differences in intrinsic properties, we only compare the results from devices that are deposited under the same conditions in one batch.

Fig.1a shows two representative voltage hysteresis loops measured across the two side electrodes from a 15 nm thick NiFe under a vertical $\nabla T$, generated by the same heater power but with opposite polarities in heater current (±80 mA). An external in-plane magnetic field is swept perpendicular to the channel. The saturation level in both loops is exactly the same, indicating that the sample is well-insulated from the heater. The horizontal loop shift (~ ±3 Oe) is clearly caused by the Oersted field produced by the heater current. In this geometry, ANE produces voltage hysteresis loops along the channel, since $V_{ANE} \propto \hat{x} \cdot (\nabla T \times \hat{m})$, where $\hat{x}$ is the unit vector along the channel (x-axis), and $\hat{m}$ is the unit vector of the magnetization which reverses the direction with hysteresis under a sweeping magnetic field (y-axis). To further verify the origin of the voltage, we apply a 400 Oe in-plane rotating field which fully saturates and rotates the magnetization direction. As the field is rotated, we observe a $\sin\theta$-dependence in the voltage (Fig. 1b). Furthermore, we find the magnitude of the voltage signal is proportional to the heater power (Figs. 1c and 1d). These facts are consistent with the properties of ANE in NiFe.

Under a fixed heater power, NiFe simply acts as a voltage source $V_N$ with an internal resistance $R_N$. When a non-magnetic layer such as Pd is deposited on top, two things will happen. First, the ANE voltage source will cause a current flow in the non-magnetic layer that reduces the voltage drop when measured in the open-circuit geometry. Second, since Pd has strong SOC, $\nabla T$ also drives a vertical spin current, $J_s$, into Pd which is converted to a charge current $J_c$ due to ISHE [7, 8]. This $\nabla T$–driven effect is the longitudinal SSE which has been extensively studied recently [1, 2, 9]. The SSE voltage, $V_S$, can be written as

$$V_S \propto R_S \theta_{SH} \lambda_S tanh\left(\frac{t}{2\lambda_S}\right), \qquad (1)$$

where $R_S$ is the resistance of the SOC layer with thickness $t$, and $\theta_{SH}$ is its spin Hall angle [10]. In the large thickness limit, i.e. $t > 2\lambda_s$, $V_s$ is simply proportional to $R_s$, or inversely proportional to the film thickness. To simplify the analysis, we stay in the thick limit for all samples, i.e. $V_s/R_s$ approaching constant. In the meantime, if the Pd is sufficiently thick, we can approximately treat it as an independent layer that is connected in parallel to the NiFe layer. Therefore, in the presence of both ANE and SSE voltage sources, the total measured voltage $V$ is proportional to the total measured resistance $R$, i.e.

$$V = \left(\frac{V_S}{R_S} + \frac{V_N}{R_N}\right) R, \qquad (2)$$

where the first and second terms in the slope represent SSE and ANE contributions, respectively. In the thick limit, the proportionality constant is independent of the SOC layer thickness. As the Pd-layer thickness is varied, the total resistance $R$ of the circuit changes, but the total voltage $V$ should remain on a straight line. This simple model predicts a slope change when Pd is placed on top of NiFe because of the extra SSE signal. However, an additional complication may arise when Pd or other SOC metal layer is in direct contact with NiFe. The proximity-induced magnetic moment in the SOC-layer experiences the same $\nabla T$ and generates its own ANE, which is difficult to be separated out and has been a subject of debate [1, 2, 4]. The presence of this proximity effect is clearly demonstrated in Fig. 2a. When a Pd layer of different thicknesses (4 and 5 nm) is placed on NiFe, $V$ is indeed directly proportional to $R$. To remove the proximity-induced ANE, we insert a thin layer of Cu with variable thicknesses (from 1 to 8 nm) between NiFe and Pd while keeping the thickness of the NiFe and Pd unchanged (both 5 nm). It is known that Cu does not produce measurable ISHE due to its small $\theta_{SH}$; furthermore, $\lambda_s$ of Cu is so long that the inserted thin Cu-layer does not effectively attenuate $J_s$ [11]. On the other hand, adding a Cu-layer reduces the total voltage signal. In the thick limit, the Cu-layer can be treated as a parallel resistor that shunts the current and Eq. 2 should also holds. Clearly, all data points fall on a straight line as shown in Fig. 2a. It suggests that the Cu-layer in this thickness range works as a load resistor. To check the validity of the parallel resistor model, we plot the total conductance as a function of the Cu-thickness and find that the relationship is approximately linear. The thinner Cu (1 nm) shows the largest deviation, indicating a possible breakdown of the model. Although the deviation for thin Cu-layers is not very obvious in the V vs. R plot, we will draw any conclusion only from the thick Cu-layers. Compared with NiFe/Cu/Pd, the slope for NiFe/Pd is larger, which clearly demonstrates the presence of the proximity-induced ANE in Pd. Additionally, the two straight lines do not merge for small Cu thicknesses, suggesting good isolation between NiFe and Pd by the Cu-layer. Fig. 2b shows the data from devices fabricated in a different batch which further confirms the role of the Cu-layer regardless of its position in the structure and the importance of the proximity effect.

In NiFe/Cu/Pd, we believe that the longitudinal SSE of Pd from $J_s$ transmitted through the Cu layer is contained in the slope. To further reveal the SSE contribution, we replace the Pd layer by a $\beta$-phase Ta layer, another strong SOC metal. It is known that the spin Hall angle $\theta_{SH}$ of the $\beta$-phase Ta is opposite to that of Pt [12]. To determine the sign of $\theta_{SH}$ of Pd relative to Ta, we have prepared YIG/Pd and YIG/Ta bilayer structures and conducted SSE measurements. Here the epitaxial YIG film (~ 50 nm thick) is grown on a gadolinium gallium garnet (110) substrate by pulsed laser deposition. The growth and magnetic properties have been reported elsewhere [13, 14]. The 5 nm thick Pd and 5 nm thick Ta are deposited in different areas of the same YIG film using sputtering. Similar heater structure to the inset of Fig. 1a is used for the longitudinal SEE measurements. We observe two opposite hysteresis loops in these two devices as shown in Fig. 3, indicating opposite signs of the spin Hall angle in Pd and Ta.

To focus on the magnitude of the SSE contribution, we compare the results in the following three sets of structures: NiFe(5 nm)/Cu(2 nm)/Pd($t_{Pd}$), NiFe (5 nm)/Cu(2 nm)/Ta($t_{Ta}$), and NiFe(5 nm)/Cu($t_{Cu}$), all prepared in the same batch to avoid any irreproducibility. In the first two sets, we only vary the thickness of Pd (3-10 nm) or Ta (10-25 nm). The third set serves as a reference that only contains a NiFe ANE voltage source and a variable Cu (2-8 nm) resistor. The first set in Fig. 4a (squares) is similar to a set shown in Fig. 2a (circles), but the difference is in the actual layer of changing thickness: Pd in Fig. 4a but Cu in Fig. 2a. In both cases, the proximity effect is eliminated by Cu. In spite of the differences in structure, both exhibit excellent linear dependence. Furthermore, even though they are fabricated in two batches, the slopes of the two lines are very close to each other, i.e. 4.92 µV/kΩ and 4.94 µV/kΩ, respectively. In the reference set (triangles), the only voltage source is the ANE voltage from the 5 nm thick NiFe, common to other structures. The slope is 3.57 µV/kΩ, clearly smaller than that of the set with varying Pd thickness. When Ta replaces Pd, we expect the straight line to lie below the line for the reference sample set due to the negative SSE signal; however, it stays above that of the reference set. Moreover, the slope of the Ta set (4.57 µV/kΩ) is only slightly small than that of the Pd set, which immediately suggests that the SSE contribution is not the primary reason for the slope enhancement observed in NiFe/Cu/Pd. The fact that both the Pd and Ta samples show significant enhancements relative to the reference indicates another more important mechanism present in both sample sets.

In the reference samples, i.e. NiFe/Cu, NiFe generates spin accumulation in Cu under a vertical $\nabla T$, but a negligible $J_s$ since Cu has a much longer $\lambda_S$ than its thickness. However, when a Pd (Ta) layer is placed on top, spin accumulation extends to Pd (Ta) on the scale of its $\lambda_S$ (~ 2 nm). It consequently draws a finite $J_s$ in Pd (Ta), which is exactly the source for the longitudinal SSE voltage from the layer. On the other hand, this $J_s$ is continuous in the Cu-layer and has to extend to NiFe. The additional $J_s$ in NiFe can generate an ISHE voltage by NiFe. The profile of spin-chemical potential, $\mu_S$, can be described by Fig. 4b. Since $J_s$ is conserved when going through the Cu-layer, our experimental results suggest that this additional ISHE voltage converted by NiFe be the source of the enhanced slope in NiFe/Cu/Pd(Ta).

Now let us estimate the spin Hall angle of NiFe. In the simple current shunting model, the ISHE voltage source contributes to the slope of the straight line by $V_S/R_S$ in Eq. 2, as discussed earlier. In fact, the total conductance vs. layer thickness shows excellent linear dependence for both Pd and Ta, justifying the parallel resistor model. Now we have two such terms, one from Pd (Ta) and the other from NiFe. Since $R_S$ for Ta is an order of magnitude larger than that of Pd, the contribution to the slope from Pd is about an order of magnitude larger than that from Ta assuming similar ISHE voltages from the two. Therefore, if we draw a line to represent the additional ISHE contribution from NiFe, this line would be just slightly above the Ta-line. With this new reference line, we can estimate the relative contributions from NiFe and Pd using the slope of these straight lines. The slope difference between the new and old reference lines is identified as $V_s/R_s$ for NiFe due to ISHE in NiFe, which is about 30% of the slope of for the ANE of NiFe. Then the slope difference between the NiFe/Cu/Pd line and the new reference line defines $V_s/R_s$ for Pd. We use Eq. 1 to relate $V_s$ from NiFe and Pd to their spin Hall angles and $\lambda_S$. We take 2 and 5 nm [15, 16] for $\lambda_S$ of Pd and NiFe, respectively, and obtain the lower bound for the ratio of the spin Hall angle between these two materials: $\frac{\theta_{SH}^{NiFe}}{\theta_{SH}^{Pd}} \sim 2.5$.

The presence of an adjacent heavy metal as a spin current drain greatly influences $J_s$ in the ferromagnet of thickness $\sim \lambda_S$, which in turn creates an additional ISHE voltage. A related effect reported by Costache et al. also illustrates the importance of the $J_s$ backflow in spin pumping devices [17]. NiFe as an ISHE detector was previously proposed in NiFe/YIG [10], where the spin current effect is superimposed on the intrinsic ANE signal of NiFe. By destroying the spin current effect through Ar-ion beam bombardment or inserting a thin MgO layer in between, the remaining effect was attributed to the intrinsic ANE of NiFe. In our NiFe/Cu/Pd(Ta) devices, the proximity-induced ANE is completely eliminated; therefore, the spin current draining effect due to the SOC metal is unambiguously singled out. Our results demonstrate the importance of the spin current manipulation in nano-scale devices.

We thank W. P. Beyermann, N. Amos, D. Humphrey, and D. Yan for their assistance. We also thank J.E. Wegrowe for useful discussions. YX, BY, and JS were supported by the DOE BES award #DE-FG02-07ER46351 and ZJ and CT were supported by DARPA/DMEA under H94003-10-2-1004.

**Figure captions:**

FIG. 1. (a) Room-temperature voltage loops of a 15 nm thick NiFe film for two heater currents ($i = \pm 80$ mA). The inset shows a schematic diagram of the device. The $Al_2O_3$ insulating layer is 300 nm thick. (b) Normalized voltage under a 400 Oe in-plane field as a function of rotation angle $\theta$ and the line is a $\sin\theta$ fit. (c) Voltage loops for heater currents ranging from 10 to 80 mA. (d) Saturation voltage vs. $i^2$, which is proportional to the heater power. The line is a linear fit.

FIG. 2. (a) Measured voltage output $V$ as a function of the total resistance $R$. The actual resistance of the Au line of a fixed length is measured and $V$ is scaled to exactly the same level of the heater power. The thickness is 5 nm for NiFe, 5 nm for Pd (except the upper-right black point at 4 nm), and 1, 3, 5 and 8 nm for Cu. (b) $V$ vs. $R$ of another batch of samples, in which NiFe, Cu and Pd thickness are all fixed at 5 nm. The insets in both (a) and (b) schematically show the device structures.

FIG. 3. Longitudinal SSE voltage loops of YIG-based structures when the same perpendicular heater power is applied. Pd and Ta layers are 5 nm thick and the devices are fabricated on the same piece of YIG.

FIG. 4. (a) Measured $V$ vs. the total resistance $R$ for three series of samples. NiFe thickness is 5 nm in all samples, and Cu thickness is 2 nm in samples represented by blue squares and green circles. Straight lines are linear fits. (b) Spin-chemical potential, $\mu_s$, in NiFe(5 nm)/Cu(2 nm) and NiFe(5 nm)/Cu(2 nm)/SOC(5 nm), where SOC represents Pd or Ta. The SOC metal behaves as a spin sink in which the spin density drops exponentially, resulting a lower $\mu_s$ in Cu. Consequently, inside NiFe, a sharper drop in $\mu_s$ creates an extra spin current that results in an ISHE voltage.

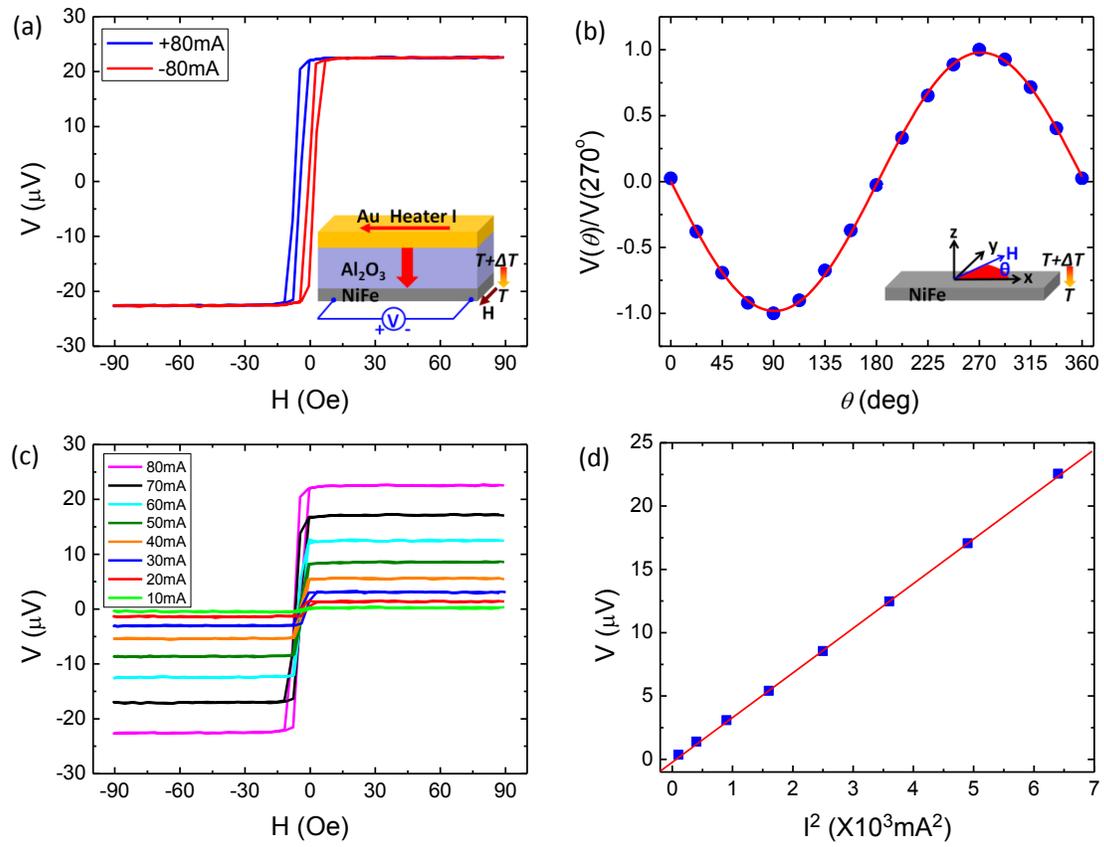

Figure 1

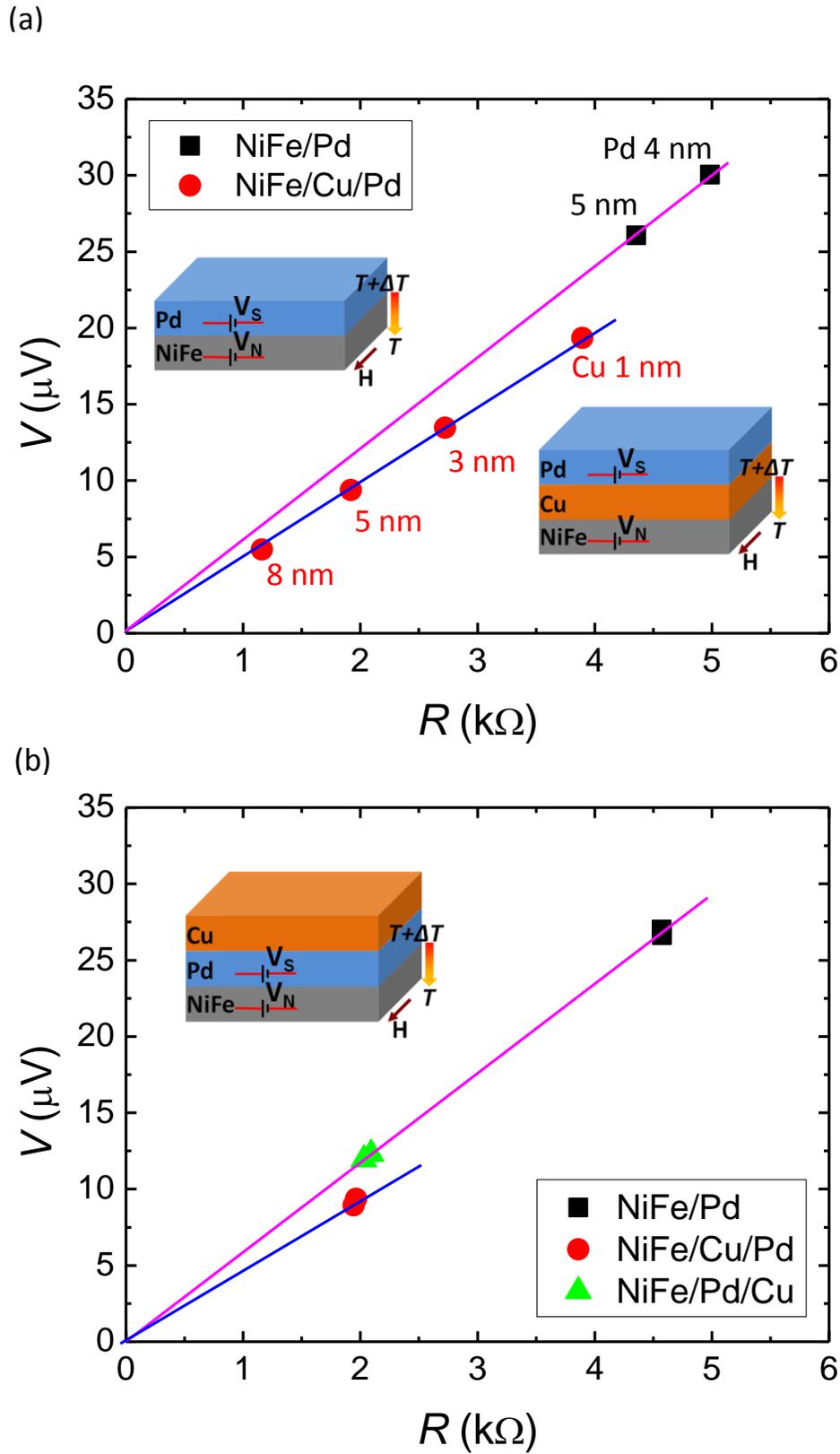

Figure 2

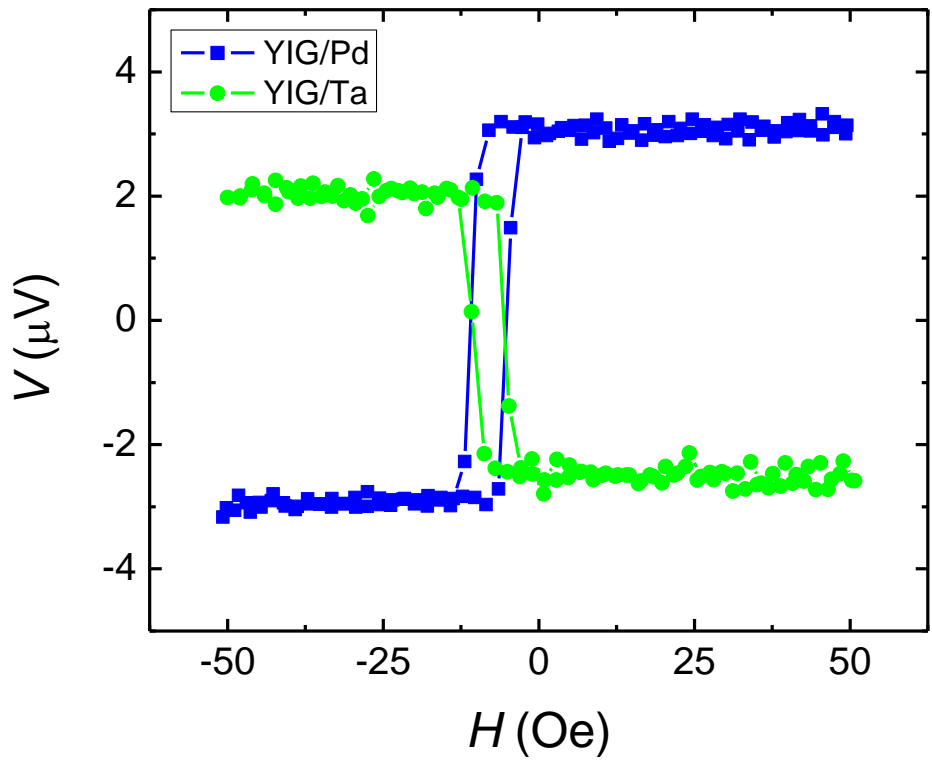

Figure 3

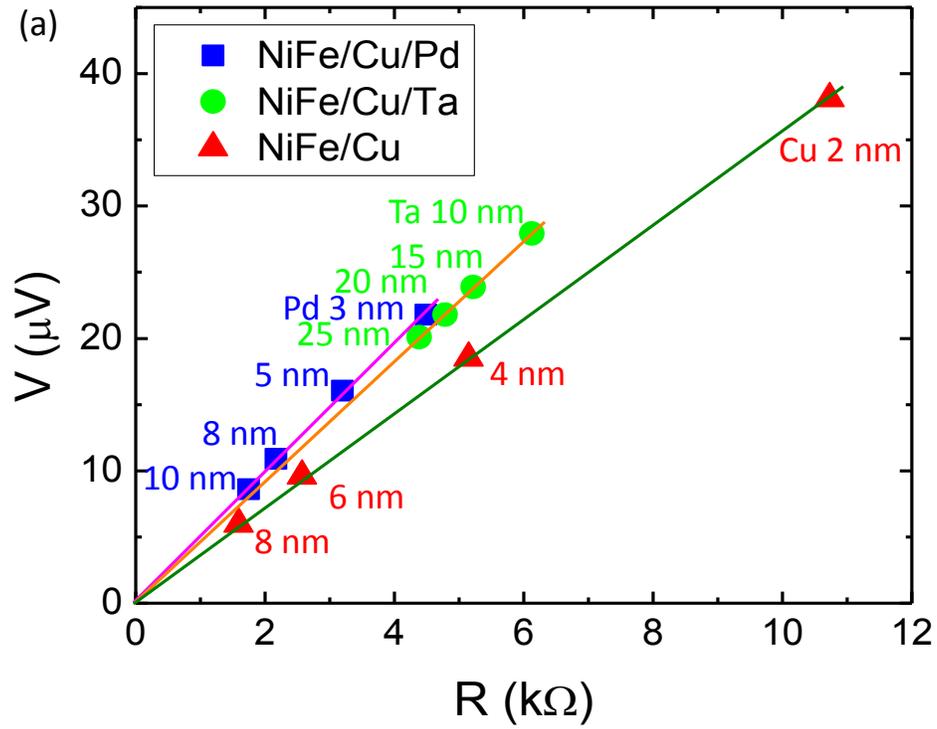

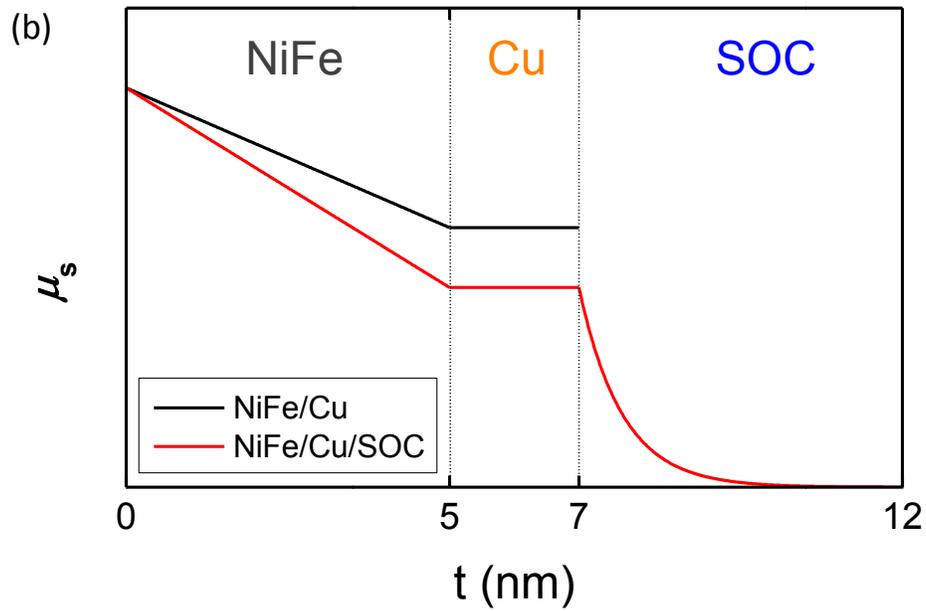

Figure 4